\documentclass[aps,eqsecnum,preprint,floats,epsf,epsfig,nofootinbib,showpacs]{revtex4}
\usepackage{graphicx}
\usepackage{amsmath}
\def\be{\begin{eqnarray}}
\def\en{\end{eqnarray}}
\def\non{\nonumber}
\def\la{\langle}
\def\ra{\rangle}
\def\pl{{ Phys. Lett.}~}
\def\pr{{ Phys. Rev.}~}

\def\bi{\bibitem}
\begin{document}

\title{\Large \bf Three symmetry breakings in strong and radiative decays of strange heavy mesons}

\author{ \bf Chi-Yee Cheung$^a$ and Chien-Wen Hwang$^b$\footnote{
t2732@nknucc.nknu.edu.tw}}
\vskip 1.4 cm
\affiliation{\centerline{$^a$ Institute of Physics, Academia Sinica, Taipei, Taiwan 115, Republic of China}\\
\centerline{$^b$ Department of Physics, National Kaohsiung Normal University, Kaohsiung,} \\
\centerline{Taiwan 824, Republic of China}}


\begin{abstract}
In this paper, we investigate three symmetry breaking effects in strong and radiative decays of strange heavy mesons. We study $1/m_Q$ corrections within the heavy quark effect theory, as well as $SU(3)$ and $SU(2)$ symmetry breakings induced by light-quark mass differences and the $\eta-\pi$ mixing vertex. These effects are studied in a covariant model. The numerical results show that the $1/m_Q$ corrections of the coupling constants are consistent with $\alpha_s \Lambda_{\textrm{QCD}}/m_Q$. The $SU(3)$ symmetry violating effect of the strong coupling constant is obviously larger than that of the magnetic coupling constant. The value of the $\eta-\pi$ mixing vertex has some changes because of the renewed data. As compared with the other theoretical calculations and the experimental data, our radiative decay rates are much larger than those of the other theoretical methods, except for $\chi$PT; however, our branching ratios are close to the experimental data.
\end{abstract}
\maketitle %
\section{Introduction}
For excited strange heavy mesons $(D^*_s,B^*_s)$, pion and/or photon emissions are the dominant decay modes which determine their lifetimes \cite{PDG14}. Of these decay modes, the radiative decay, $D^*_s\to D_s\gamma$, and the only kinematically allowed strong decay, $D^*_s\to D_s\pi$, which is the isospin-violating mode, have been observed, and the branching ratio $\Gamma(D^*_s\to D_s\pi)/\Gamma(D^*_s\to D_s\gamma)$ has been measured by the CLEO \cite{exp1995} and BaBar collaborations \cite{exp2005}. The latter collaboration obtained $\Gamma(D^*_s\to D_s\pi)/\Gamma(D^*_s\to D_s\gamma)=0.062\pm 0.005(\textrm{stat.})\pm 0.006(\textrm{syst.})$, which was a significant improvement over the former one. This precise value provides an ideal occasion to test different theoretical estimations for the strong and electromagnetic interactions of strange heavy mesons.

In 1989, it was realized that, in low-energy situations where the typical gluon momenta are small compared with the heavy quark mass $(m_Q)$, quantum chromodynamics (QCD) dynamics becomes independent of the flavor (mass) and spin of the heavy quark \cite{iw,iw2,Geo1}. These new spin and flavor symmetries combine to form an $SU(2N^Q_f)$ symmetry, called heavy quark symmetry (HQS), which is not manifest in the original QCD Lagrangian. 
HQS allows us to factorize the complicated light quark and gluon dynamics from that of the heavy one, and thus provides a clearer physical picture in the study of heavy quark physics. Beyond the symmetry limit, a heavy quark effective theory (HQET) can be developed by systematically expanding the QCD Lagrangian in powers of $1/m_Q$, with which HQS breaking effects can be studied order by order \cite{Geo1,NN,MRR}. Although the development of HQET from QCD has simplified the analysis of heavy hadron physics, many properties of hadrons, for example, their decay constants and axial coupling constants, are still not calculable directly from QCD. To study these quantities, one unavoidably has to use phenomenological models to describe the structures of hadrons. These include the constituent quark model (CQM) \cite{ISGW,ISGW2}, the MIT bag model \cite{SZ,HwangMIT}, the lattice QCD calculations \cite{lQCD1,lQCD2}, QCD sum rules \cite{BBK}, and the light-front quark model (LFQM) \cite{Jaus,Jaus1,Jaus2}. In spite of the fact that the CQM and the MIT bag models have been widely used, the results calculated from these two models are trustworthy only for processes involving small momentum transfers. The LFQM is a relativistic quark model with
simple boost kinematics which allows us to describe physical processes with large momentum transfers. However, this model is not a fully Lorentz covariant \cite{CCHZ}, and this defect limits its usefulness to  matrix elements with space-like momentum transfers ($q^2 \leq 0$) only. Moreover, the LFQM is not
capable of handling the so-called Z-diagrams \cite{CCH}. In Ref. \cite{CCHZ}, a covariant light-front model of heavy mesons has been suggested. However, the approach taken there is not systematic, and light-quark currents are not considered. To overcome the drawbacks mentioned above, a covariant field theoretical model has been proposed for the heavy meson bound state problem \cite{CZ,CCZ,CZ1}. This model is fully covariant and satisfies HQS; at the same time, it retains the simplicity of the quark model picture. This theory allows us to formulate theoretical calculations in terms of the standard Feynman diagrams.
Therefore, the lack of Z-diagrams in the ordinary LFQM is no longer a problem. Combining this model with HQET, we can systematically study various $1/m_Q$ corrections to heavy meson properties in the framework of perturbative field theory.

In the other extreme, due to the relatively small light-quark masses $(m_u, m_d, m_s)$, the light-quark sector of the QCD Lagrangian obeys an approximate $SU(3)_L \times SU(3)_R$ chiral symmetry \cite{DGH}.
Due to the spontaneous breaking of the chiral symmetry, there exist eight pseudoscalar bosons (called Goldstone bosons, which include three $\pi$'s, four $K$'s, and one $\eta$), whose dynamics obeys the $SU(3)_L \times SU(3)_R$ chiral symmetry. 
If we want to study the low-energy interactions of heavy hadrons and Goldstone bosons, we need to build an effective theory that obeys both chiral and heavy quark symmetries. This was done in references \cite{Wise92,BD92,HYC1,HYC3,HYC4}, where chiral symmetry and HQS were synthesized in a single effective chiral Lagrangian which described the strong interactions between heavy hadrons and Goldstone bosons. The theory has since been extended to incorporate electromagnetic interactions as well \cite{CG,Amundson,HYC2,HYC3,HYC4}.
In principle, the effective chiral Lagrangian provides an ideal framework in which to study the strong decay mode. However, symmetry considerations alone, in general, do not lead to quantitative predictions, unless further assumptions are made to extract the values of the various coupling constants appearing in the Lagrangian. Furthermore, the framework of an effective chiral Lagrangian does not allow for a systematic discussion of HQS violating $1/m_Q$ effects, which is important for a thorough understanding of heavy quark physics. In fact, in
the heavy-light $(Q\bar s)$ system, there are three different types of symmetry breaking mechanisms: (1) HQS breaking from $1/m_Q$ corrections, (2) $SU(3)$ symmetry breaking due to strange quark mass $(m_s \not= m_{u,d})$, and (3) $SU(2)$ symmetry breaking due to the up-down quark mass difference $(m_u \not= m_d)$. 
The purpose of this paper is to systematically study these symmetry breaking effects in a covariant model for the strong and radiative decays of strange heavy mesons.

The paper is organized as follows. 
In Sec. II, we briefly review the covariant model, which is based on HQET. Some heavy meson properties in the heavy quark limit and $1/m_Q$ corrections are considered. The numerical calculations and discussions are expressed in Sec. III. In Sec. IV, we make some concluding remarks.


\section{Formalism}
The covariant model starts from HQET in the heavy quark limit $(m_Q \to \infty)$ and describes a heavy meson as a composite particle, consisting of a reduced heavy quark coupled with a brown muck of light degrees of freedom. It is formulated in an effective Lagrangian approach, so that it is fully covariant, and we can use Feynman diagrammatic techniques to evaluate various processes.

\subsection{Covariant model}
Using the $1/m_Q$ expansion to the heavy quark QCD Lagrangian \cite{Geo1,NN}, the QCD Lagrangian for heavy and light quarks plus gluons can be written as $L=L_0+L_{m_Q}$, where
 \be
 L_0&=& \bar h_v iv\cdot D h_v + \bar q~(i\gamma_\mu D^\mu - m_q)~q - \frac{1}{4} F^{\mu\nu}_a F_{a\mu\nu}, \label{L0}\\
 L_{m_Q}&=&\sum_{n=1}^{\infty}\left(\frac{1}{2 m_Q}\right)^n\bar h_v i \not\!\!D_{\bot} (-i v \cdot\! D)^{n-1} i
 \not\!\!D_\bot h_v, \label{expand}
 \en
$D^\mu_{\bot} = D^\mu - v^\mu v \cdot D$ is orthogonal to the heavy quark velocity, $L_0$ is responsible for binding a heavy quark and a light quark in the heavy quark limit, and $L_{m_Q}$ contains
$1/m_Q$ corrections to $L_0$. The effective Lagrangian we have constructed to describe the low-energy dynamics of pseudoscalar heavy mesons reads \cite{CZ,CCZ,CZ1}:
 \be
 L_{\textrm{eff}}=L+\Phi_v^\dag(iv\cdot \stackrel{\leftrightarrow}{\partial}-2\bar \Lambda)\Phi_v-\bar h_v i\gamma_5 q_v \Phi_v +\textrm{h.c.}
 \en
where $\Phi_v$ represent the composite pseudoscalar heavy meson fields which appear only as external states,
 \be
 \bar \Lambda \equiv \lim_{m_Q \to \infty} m_M-m_Q \label{Lambda}
 \en
is their residual mass in the heavy quark limit,
 \be
 q_v=G F(-iv\cdot \partial) q
 \en
represents collectively the degrees of freedom in a heavy meson,
where $F$ is a form factor whose presence is expected for an effective interaction resulting from the non-perturbative QCD dynamics, and $G$ is the normalization constant given by
 \be
   G^{-2} = i \int \frac{d^4 p}{(2 \pi)^4} F^2(v\cdot p) \frac{v\cdot p + m_q}{(\bar \Lambda-v\cdot p)^2(p^2-m^2_q)}. \label{Gno}
 \en
At this point, we note that $F(v\cdot p)$ is analogous to the meson wave function in the LFQM, and $G$ is the corresponding normalization constant. To explicitly evaluate $G$ and other physical quantities, we need to specify the structure function $F(v\cdot p)$, which is unfortunately not calculable from first principles.  Nevertheless, from the constraints that $F$ does not depend on the heavy quark residual momentum and it forbids on-shell dissociation of the heavy meson into $Q\bar q$, a plausible form for $F$ is:
 \be
 F(v\cdot p) = \varphi(v\cdot p) (\bar \Lambda-v\cdot p), \label{FF}
 \en
where the function $\varphi(v\cdot p)$ does not have a pole at $v\cdot p = \bar \Lambda$.

\begin{figure}
 \includegraphics*[width=5in]{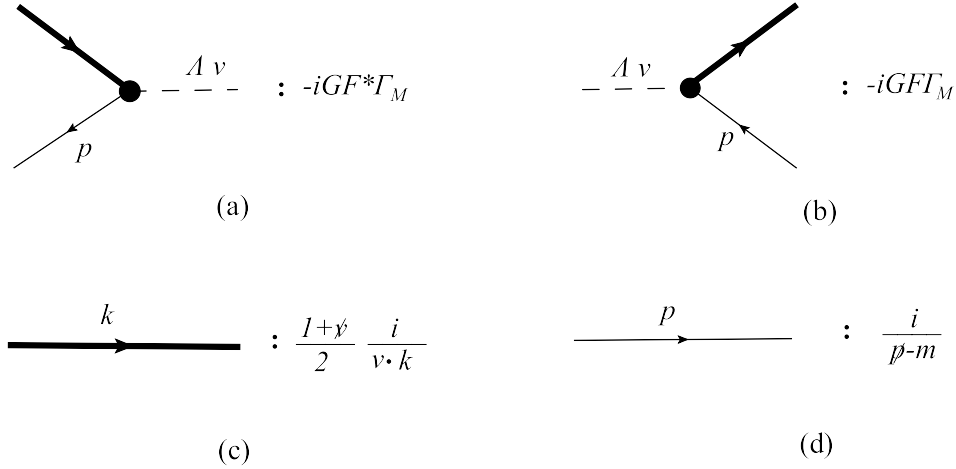}
 \caption{Feynman rules in the heavy quark limit.}
  \label{fig:2}
 \end{figure}

Within this framework, hadronic matrix elements are calculated via standard Feynman diagrams where heavy mesons always appear as external legs. The Feynman rules for this effective theory are shown in Figure \ref{fig:2}. Figure \ref{fig:2}(a) specifies the meson-$Q$-$q$ vertex with $\Gamma_M = i\gamma_5 (-\not\!\epsilon)$, for $M$ is the pseudoscalar (vector) meson. All the other Feynman rules are the same as in QCD and HQET.
Thus, if the power of $|\vec{p}|$ in the wave function $\varphi$ is less than $-\frac{2}{3}$, this model will work well. After building a covariant framework to describe heavy meson structures, we go on to evaluate some of the basic heavy meson properties. These include the decay constant, the $1/m_Q$ corrections of the heavy meson mass, and the axial-vector and electromagnetic coupling constants of strange heavy mesons.
\subsection{Decay constants and $1/m_Q$ corrections of the heavy meson mass}
\begin{figure}
 \includegraphics*[width=2.5in]{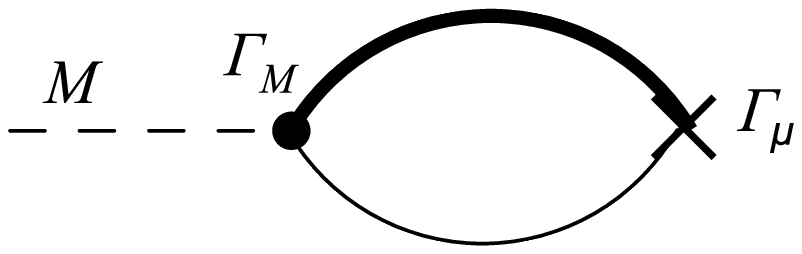}
 \caption{Feynman diagram for heavy meson decay constant.}
  \label{fig:dc}
   \end{figure}

Consider the heavy meson decay constants defined by:
 \be
   \la 0| \bar q \gamma^\mu \gamma_5 h_v|P(v)\ra &=& i\bar f_P v^\mu, \non \\
   \la 0| \bar q \gamma^\mu h_v|V(\epsilon)\ra &=& \bar f_V \epsilon^\mu. \non
 \en
The Feynman diagram to be evaluated is illustrated in Figure \ref{fig:dc}.
Using the Feynman rules in Figure \ref{fig:2}, the matrix element is evaluated as:
 \be
   \langle 0|\bar \psi_q \Gamma_\mu h_v|M(v)\rangle 
 &=& 2\sqrt{N_c}i \int \frac{d^4p}{(2\pi)^4} \frac{GF(v\cdot p)(v\cdot p+m_q)}{(p^2-m^2_q+i\varepsilon)(\bar \Lambda -v\cdot p+i\varepsilon)}Tr\Bigg[\frac{-1}{4}\Gamma_\mu (1+\not\! v)\Gamma_M\Bigg] \non \\
  &\equiv& \bar f_M~\textrm{Tr}\Bigg[\frac{-1}{4}\Gamma_\mu (1+\not\! v)\Gamma_M\Bigg],\label{dch}
 \en
where $N_c=3$ is the number of colors, while $\sqrt{N_c}$ arises from the color wave function of meson, $(r\bar r+g\bar g+b\bar b)/\sqrt{N_c}$,
and $\Gamma_M = i\gamma_5 (-\not\!\!\epsilon)$ for a pseudoscalar (vector) heavy meson; the corresponding weak current vertex is $\Gamma_\mu = \gamma_\mu\gamma_5 (\gamma_\mu)$. Here, as mentioned in the last subsection, the meson field is represented by the form factor $F$. Thus, the decay constant in the heavy quark limit is given by:
 \be
   \bar f_M = 2\sqrt{3}iG \int \frac{d^4p}{(2\pi)^4} \frac{\varphi(v\cdot p)(v\cdot p+m_q)}{(p^2-m^2_q)}.\label{fbm}
 \en 
We find that this decay constant is the same for pseudoscalar and
vector heavy mesons, which is in accord with the prediction of HQS.
$\bar f_M$ is related to the usual definition of decay constant
$f_M$ by $f_M = \bar f_M/\sqrt{m_M}$.

\begin{figure}
 \includegraphics*[width=6in]{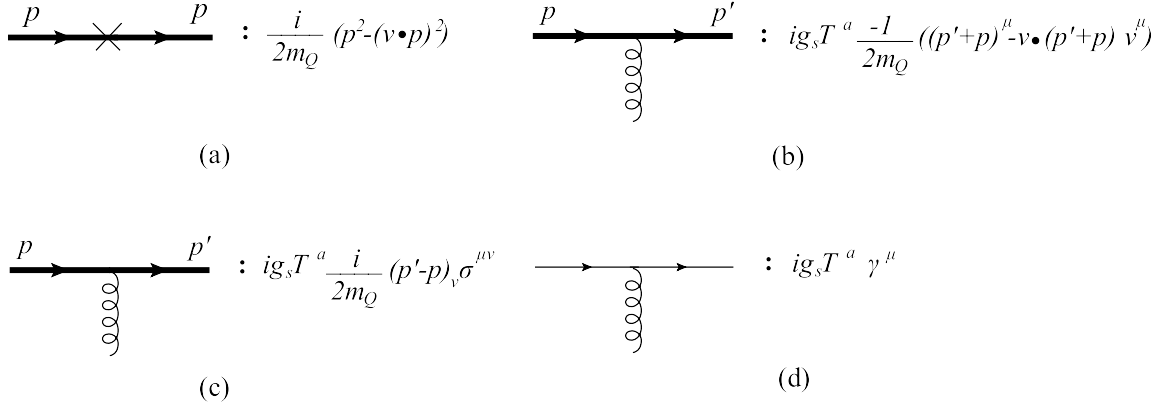}
 \caption{(a), (b), and (c) are Feynman rules for ${\cal O}^{(k)}_1$, ${\cal O}^{(g)}_1$,
 and ${\cal O}^{(g)}_2$; (d) is the light quark coupling to a gluon.}
  \label{fig:3}
 \end{figure}

\begin{figure}
 \includegraphics*[width=5in]{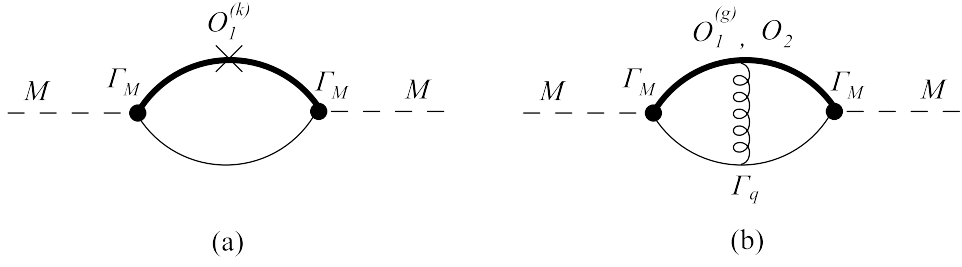}
 \caption{Feynman diagrams for $1/m_Q$ corrections to meson mass.}
  \label{fig:4}
 \end{figure}

Next, we consider the $1/m_Q$ corrections of the heavy meson mass. The Lagrangian in Eq. (\ref{expand}) can be expanded as:
\be
L_{m_Q}={\cal O}_1 + {\cal O}_2 + {\cal O} \left(\frac{1}{m^2_Q}\right),\label{O12}
\en
where ${\cal O}_1 = \frac{1}{2 m_Q}~{\bar h}_v~(iD_\bot)^2~ h_v$,  
${\cal O}_2 = \frac{g}{4 m_Q}~{\bar h}_v~\sigma^{\mu\nu}~G_{\mu\nu}~h_v$, 
and $G^{\alpha\beta} = T_a G^{\alpha\beta}_a = \frac{i}{g_s}[D^\alpha,D^\beta]$ is the gluon field strength tensor. ${\cal O}_1$ is the gauge invariant extension of the kinetic energy arising from the off-shell residual motion of the heavy quark, and ${\cal O}_2$ describes the color magnetic interaction of the heavy quark spin with the gluon field. It is clear that both ${\cal O}_1$ and ${\cal O}_2$ break the flavor symmetry, while ${\cal O}_2$ breaks the spin symmetry.
The Feynman rules for these HQS breaking interactions are given in Figure \ref{fig:3}.
With the $1/m_Q$ corrections included, the heavy meson masses can be expressed as:
 \be
   m_M = m_Q + \bar \Lambda -\frac{1}{2 m_Q}(\lambda_1 + d_M \lambda_2),\label{MMM}
 \en
where $d_M=3(-1)$ for the pseudoscalar (vector) meson, $\lambda_1$ comes from ${\cal O}_1$, and $\lambda_2$ comes
from ${\cal O}_2$. $\lambda_1$ receives two different contributions, which are a kinetic energy piece and a one-gluon exchange piece,
thus, $\lambda_1 = \lambda_1^{(k)} + \lambda_1^{(g)}$.
The relevant Feynman diagrams are shown in Figure \ref{fig:4}.
Using Feynman rules in Figures \ref{fig:2} and \ref{fig:3}, we can readily write down the various contributions:
 \be
    \lambda^{(k)}_1 
    &=& i G^2\int \frac{d^4 p}{(2\pi)^4} \frac{|\varphi(v\cdot p)|^2} {(p^2-m^2_q)} 2(p^2-v\cdot p^2) (v\cdot p +m_q), \label{lamonek} \\
     \lambda^{(g)}_1 
    &=&-C_f G^2 g^2_s \int \frac{d^4 p~d^4 p'}{(2\pi)^4(2\pi)^4} \frac{\varphi^\dagger(v\cdot p') \varphi(v\cdot p)}{(p'^2-m^2_q) (p^2-m^2_q) (p-p')^2}~{\cal T}^1_M ,\label{lamone} \\
   d_M \lambda_2 
  &=& -g_s^2 C_f G^2\int \frac{d^4 p~d^4 p'}{(2\pi)^4(2\pi)^4} \frac{\varphi^\dagger(v\cdot p') \varphi(v\cdot p)}{(p'^2-m^2_q) (p^2-m^2_q) (p-p')^2}~{\cal T}^2_M \label{lamtwo}
 \en
where $C_f=\frac{4}{3}$ is a color factor and ${\cal T}^{1,2}_M$ 
are defined by:
 \be
   {\cal T}^1_M &\equiv& 2\{ (p\cdot p'+p'^2-v\cdot p v\cdot p'-v\cdot p'^2)(m_q+v\cdot p) \non \\
 &&+(p\cdot p'+p^2-v\cdot p v\cdot p'-v\cdot p^2)(m_q+v\cdot p')\}, \\
   {\cal T}^2_M &\equiv& \frac{4}{3}d_M\{(p'^2-p\cdot p'+v\cdot p v\cdot p'-v\cdot p'^2)(m_q+v\cdot p) \non \\
 &&-(p\cdot p'-p^2-v\cdot p v\cdot p'+v\cdot p^2)(m_q+v\cdot p')\}.
 \en
As expected, $\lambda_1^{(k)}$ and $\lambda^{(g)}_1$ are the same for both pseudoscalar and vector mesons.
The hyperfine mass splitting is obtained :
\be
   \Delta m_{_{HF}} = m_V -m_P = \frac{2 \lambda_2}{m_Q}. \label{masslp}
\en
\subsection{Strong coupling constant}
\begin{figure}
 \includegraphics*[width=2.5in]{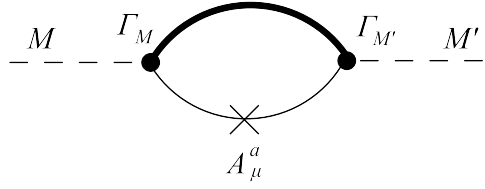}
 \caption{Feynman diagram of $f_0$.}
  \label{fig:fg0}
 \end{figure}

\def\pke%
{\begin{array}{ccc}
\frac{\pi^0}{\sqrt{2}} + \frac{\eta}{\sqrt{6}} & \pi^+ & K^+ \\
\pi^- & -\frac{\pi^0}{\sqrt{2}} + \frac{\eta}{\sqrt{6}} & K^0 \\
K^- & \bar K^0 & -\sqrt {\frac{2}{3}} \eta
\end{array}}
First, we study the zero order of strong coupling constants. An effective Lagrangian of pseudoscalar ($P$) and vector ($V$) mesons and their couplings to the Goldstone bosons is constructed as \cite {HYC1}:
\be
   {\cal L}_{VP} &=& {\cal D}_\mu P~{\cal D}^\mu P^\dagger - M^2_H PP^\dagger + if M_H (P~{\cal A}^\mu V_\mu^\dagger - V_\mu {\cal A}^\mu P^\dagger)-\frac{1}{2} V^{\mu\nu}V^\dagger_{\mu\nu}  \non \\
&+& M^2_H V^\mu V_\mu^\dagger+\frac{1}{2}g \epsilon _{\mu\nu\alpha\beta}(V^{\mu\nu}{\cal A}^\alpha V^{\beta\dagger} + V^\beta {\cal A}^\alpha V^{\mu\nu\dagger}), \label{LVP}
\en
where ${\cal D}_\mu P^\dagger \equiv (\partial_\mu + {\cal V}_\mu)P^\dagger$, $V^\dagger_{\mu\nu} = {\cal D}_\mu V_\nu^\dagger - {\cal D}_\nu V_\mu^\dagger$, and ${\cal V}_\mu ({\cal A}_\mu)$ is the (axial) vector field:
\be
{\cal V}_\mu &=& \frac{1}{2}(\xi^\dagger \partial_\mu \xi + \xi \partial_\mu \xi^\dagger), \label{vector}\\
   {\cal A}_\mu &=& \frac{i}{2}(\xi^\dagger \partial_\mu \xi - \xi \partial_\mu \xi^\dagger). \label{axial}
\en
$\xi$ is defined as $\xi\equiv e^{iM/f_\pi}$, $M$ is a $3 \times 3$ matrix for the octet of Goldstone bosons:
\be
  M= \left[
     \pke
     \right]
\en
and $f_\pi$ is the pion decay constant. 
Through the partial conservation of axial-vector current (PCAC), a soft pion amplitude can be related to a matrix element of the axial-vector current $A^a_\mu = \bar \psi
\frac{\lambda^a}{2} \gamma_\mu \gamma_5 \psi$ as:
 \be
   \la B \pi^a (q) |A\ra = \frac{q^\mu}{f_\pi} \la B |A^a_\mu | A\ra, \label{PCAC1}
 \en
From the chiral Lagrangian, we obtain:
 \be
   \la P \pi^a (q) |V \ra = \frac{-i}{f_\pi}\frac{f}{2}~q \cdot \epsilon.  
 \en
On the other hand, the matrix element on the right hand side of Eq. (\ref{PCAC1}) can be evaluated in the covariant model.
The Feynman diagram to be evaluated is illustrated in Figure \ref{fig:fg0}, and the relevant Feynman rules 
are illustrated in Figure \ref{fig:2}.
The result is:
 \be
   \langle M'(v)|\bar \psi_q T^a\gamma_\mu \gamma_5 \psi_q|M(v)\rangle 
 \equiv {\cal
 G}~\textrm{Tr}\Bigg[\gamma_\mu\gamma_5\Gamma_{M'}\frac{(1+\not\! v)}
 {4}\Gamma_M\Bigg]\chi^\dagger_{_{M'}} \lambda^a \chi_{_M},
 \label{fg0}
 \en
where the $\chi$ are $SU(3)$ wave functions of the heavy mesons and
 \be
   {\cal G} = \frac{-i}{3}G^2 \int \frac{d^4p}{(2\pi)^4} |\varphi(v\cdot p)|^2 (\bar \Lambda -v\cdot p) \frac{3m^2_q+p^2+2(v\cdot p)^2+6m_q v\cdot p} {(p^2-m^2_q)^2}. \label{gg}
 \en
For $V \to P\pi~(\Gamma_M = -\not\!\epsilon,~\Gamma_{M'}= i\gamma_5)$, we compare Eq. (\ref{fg0}) with Eq. (\ref{LVP}) and
where the subscript $0$ denotes zeroth order in $1/m_Q$.

\begin{figure}
 \includegraphics*[width=5in]{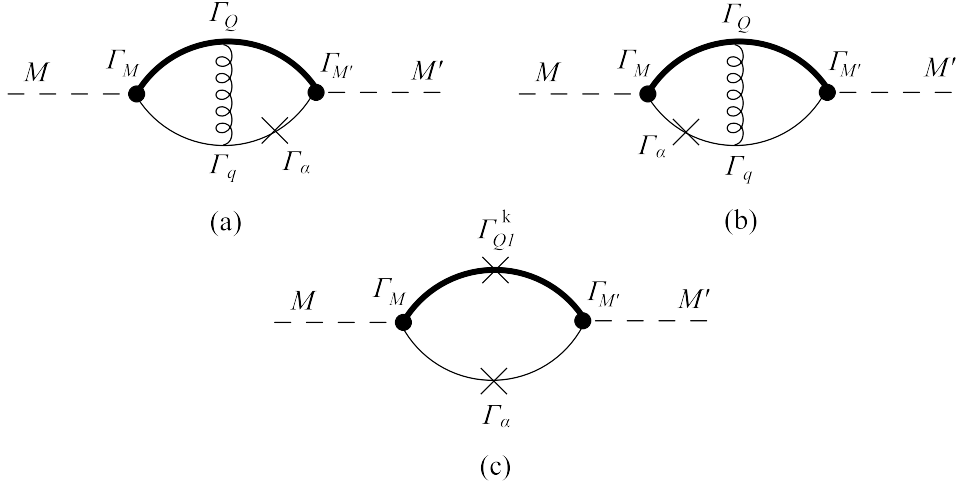}
 \caption{$1/m_Q$ corrections to strong $f$ and magnetic $d$ coupling constants. $\Gamma_\alpha$ stands for an external current. Other notations are defined in Figure \ref{fig:3}.}
  \label{fig:5}
\end{figure}

Next, we shall calculate the first order $1/m_Q$ corrections of strong coupling constants. The relevant matrix elements are
collectively illustrated in Figure \ref{fig:5}.
For $V\to P \pi$ and $\Gamma_Q=\Gamma^{(g)}_{Q1}$, we can readily write down the
matrix elements and calculate the traces for Figures \ref{fig:5} (a) as:
 \be
   {\cal M}^{(a)}_\alpha &=& \frac{C_f G^2}{2m_Q} g^2_s\int \frac{d^4 p~d^4 p'}{(2\pi)^4(2\pi)^4} \frac{\varphi(v\cdot p) \varphi(v\cdot p')}{(p^2-m^2_q)(p'^2-m^2_q)^2(p'-p)^2} {\cal R}_1^{(a)}\label{dfga}
 \en
where
 \be
   {\cal R}_1^{(a)} &=& \Bigg\{(p'\cdot p+p^2-v\cdot p'~v\cdot p - (v\cdot p)^2)\left[(m_q + v\cdot p')^2+\frac{1}{3}(p'^2-v\cdot p'^2)\right]\non \\
 &&~~+(p'^2+p'\cdot p-(v\cdot p')^2-v\cdot p' v\cdot p)\frac{2}{3}(m_q+ v\cdot p)(2m_q +v\cdot p')\Bigg\}.
 \en
Similarly, we can evaluate ${\cal M}^{(b)}_\alpha$ for Figure \ref{fig:5} (b), and it turns out that ${\cal M}^{(a)}_\alpha = {\cal M}^{(b)}_\alpha$. Then a comparison with the chiral Lagrangian result shows:
 \be
   \delta f^{(g)}_1 &=& -2\frac{C_f G^2}{ m_Q} \int \frac{d^4 p~d^4 p'} {(2\pi)^4(2\pi)^4}
   \frac{g^2_s\varphi^\dag(v\cdot p') \varphi(v\cdot p)}
   {(p^2-m^2_q)(p'^2-m^2_q)^2(p'-p)^2} {\cal R}^{(a)}_1.\label{g1g}
 \en
The above calculation can be repeated for $\Gamma_Q = \Gamma_{Q2}$.
We find that $\delta f_2$ is given by:
 \be
 \delta f_2 = \frac{C_fG^2}{m_Q} \int \frac{d^4 p~d^4 p'}{(2\pi)^4(2\pi)^4}
 \frac{g^2_s\varphi^\dag(v\cdot p') \varphi(v\cdot p)}{(p^2-m^2_q)(p'^2-m^2_q)^2(p'-p)^2} {\cal R}^{(a)}_2,\label{g2}
 \label{g2a}
 \en
where
 \be
   {\cal R}_2^{(a)} = \left(\frac{2}{3}\right)\Bigg\{\Bigg[p'^2-p'\cdot p-(v\cdot p')^2+v\cdot p' v\cdot p
   \Bigg](2m^2_q+2m_q v\cdot p +v\cdot p'~v\cdot p-p'\cdot p)\non \\
 -\Bigg[p'\cdot p-p^2-v\cdot p'~v\cdot p + (v\cdot p)^2\Bigg](m^2_q +2 m_q v\cdot p'+(v\cdot p')^2)\Bigg\}.~~~~~~~~~~~
 \en
Figure \ref{fig:5} (c) corresponds to the contribution from the heavy quark kinetic energy. For $V \to P\pi$, the matrix element
can be simplified as:
 \be
  {\cal M}^{(c)}_\alpha 
 = \delta f^{(k)}_1 \textrm{Tr}\Bigg[\gamma_\alpha\gamma_5(-\not \!\epsilon')\frac{(1+\not\! v)}{4}(-\not \!\epsilon)\Bigg]
 \chi^\dagger_{_{M'}} \lambda^a \chi_{_M},
 \en
where
 \be
   \delta f^{(k)}_1 = 2i G^2 \int \frac{d^4p}{(2\pi)^4}\frac{|\varphi(v\cdot p)|^2}{(p^2-m^2+i\epsilon)^2}
  \frac{(p^2-v\cdot p^2)}{2m_Q} \left[(m+v\cdot p)^2+\frac{1}{3}(p^2-v\cdot p^2)\right].
 \label{g11k}
 \en
Therefore, we obtain the strong coupling constant, including the $1/m_Q$ corrections, as:
 \be
   f&=& f_0 + \delta f_1^{(k)} + \delta f_1^{(g)} + \delta f_2.
 \en
\subsection{Magnetic coupling constant}
\def\udsq%
{\begin{array}{ccc}
\frac{2}{3} & 0 & 0 \\
0 & \frac{-1}{3} & 0 \\
0 & 0 & \frac{-1}{3}
\end{array}}
We now consider the coupling constant 
which governs the decay $V \to P \gamma$.
The relevant lowest-order chiral and gauge-invariant Lagrangian is given by \cite{HYC2}:
\be
   {\cal L}'_{VP} = M_H~\epsilon_{\mu\nu\alpha\beta}v^\alpha V^\beta \times \left[\frac{1}{2}d(\xi^\dagger {\cal Q}\xi+\xi {\cal Q}\xi^\dagger) \right]F^{\mu\nu}P^\dagger + {\rm h.c.}, \label{4d}
\en
where
 \be
  {\cal Q}= \left[
     \udsq
     \right],
 \en
is the light-quark charge. In the $m_Q \to \infty$ limit, the Feynman diagram to be calculated is similar to Figure \ref{fig:fg0}, except that the axial-vector current $A^a_\mu$ is replaced by the light-quark electromagnetic current $j_\mu = ee_q \bar \psi_q \gamma_\mu \psi_q$. The result is:
 \be
   \langle M'(v)|\bar \psi_q (iee_q\gamma_\mu) \psi_q|M(v)\rangle 
 \equiv {\cal D}e_q\textrm{Tr}\left[i\gamma_\mu\not\! q\Gamma_{M'}\frac{1+\not\! v}{4}\Gamma_M \right], \label{d10}
 \en
where 
 \be
 {\cal D} = 2i e G^2 \int \frac{d^4p}{(2\pi)^4} |\varphi(v\cdot p)|^2(\bar \Lambda -v\cdot p)\frac{v\cdot p+m_q}{(p^2-m^2_q)^2}, \label{D11}
 \en
and $q=p'-p \to 0$. For $V \to P \gamma ~(\Gamma_M = -\not\!\epsilon,~\Gamma_{M'}= i\gamma_5)$, we compare Eq. (\ref{d10}) with Eq. (\ref{4d}) and obtain
$d_0 = {\cal D}/2$.

Next, we calculate $1/m_Q$ corrections to the magnetic coupling $d$ corresponding to $V \to P\gamma$. The relevant Feynman diagrams are shown in Figure \ref{fig:5} with $\Gamma_\alpha = iee_q \gamma_\alpha$. For $V\to P \gamma$,
 \be
 \langle P \gamma (q,\varepsilon)|V(\epsilon)\rangle = ie_q~2d~\epsilon_{\mu\nu\alpha\beta}\varepsilon^\mu q^\nu v^\alpha \epsilon^\beta, \label{vpst1}
 \en
which comes from the effective chiral Lagrangian Eq. (\ref{4d}), the calculated procedures are similar to those of the strong coupling constants. Here, we only show the results for $\Gamma_Q = \Gamma^{(g)}_{Q1}$,
 \be
 \delta d^{(g)}_1 &=& -\frac{2g^2_sG^2}{3m_Q}\int \frac{d^4 p~d^4 p'}{(2\pi)^8} \frac{\varphi^\dag(v\cdot p')\varphi(v\cdot p){\cal S}_1^{(a)}}{(p^2-m^2_q)(p'^2-m^2_q)^2(p'-p)^2},
 \label{dggg1}
 \en
where
 \be
  {\cal S}^{(a)}_{1} &=& -2\Bigg\{\Bigg[p'^2+p'\cdot p-(v\cdot p')^2-v\cdot p' v\cdot p\Bigg](v\cdot p +m_q) ~~\non \\
 &&~~~+\Bigg[p'\cdot p+p^2-v\cdot p'~v\cdot p - (v\cdot p)^2\Bigg](v\cdot p'+m_q) \Bigg\},
 \en
and for $\Gamma_Q = \Gamma^{(g)}_{Q2}$,
 \be
 \delta d_2 &=& \frac{2g^2_sG^2}{3m_Q}\int \frac{d^4 p~d^4 p'}{(2\pi)^8} \frac{\varphi^\dag(v\cdot p')\varphi(v\cdot p){\cal S}_2^{(a)}}{(p^2-m^2_q)(p'^2-m^2_q)^2(p'-p)^2},
 \label{dggg2}
 \en
where
 \be
  {\cal S}^{(a)}_2 &=& \frac{4}{3}\Bigg\{\Bigg[p'^2-p'\cdot p-(v\cdot p')^2+v\cdot p' v\cdot p\Bigg](v\cdot p +m_q) \non \\
 &&-\Bigg[p'\cdot p-p^2-v\cdot p'~v\cdot p + (v\cdot p)^2\Bigg](v\cdot p'+m_q) \Bigg\}.
 \en
For Figure \ref{fig:5} (c), we obtain:
\be
   \delta d_1^{(k)} =-ieG^2\int\frac{ d^4p}{(2 \pi)^4} \frac{\varphi(v\cdot p)^2 (v\cdot p+m_q) (p^2-v\cdot p^2)}{2 m_Q(p^2-m_q^2)^2}. \label{dkk1}
\en
In radiative decay, there is an additional $1/m_Q$ correction which comes from the magnetic moment of the heavy quark. The matrix element of this process is:
 \be
   \langle P|\bar \psi_Q \frac{i^2ee_Q}{2 m_Q}\sigma_{\mu\nu}q^\nu \psi_Q|V(\epsilon)\rangle
 =\frac{i e}{m_Q} G^2\int \frac{d^4p}{(2\pi)^4} \frac{\varphi(v\cdot p)^2} {(p^2-m^2_q)}(v\cdot p +m_q) ie_Q \epsilon_{\mu\nu\alpha\beta} q^\nu v^\alpha \epsilon^\beta, \label{firstdd}
 \en
if $\Gamma_{M'} = i\gamma_5, \Gamma_{M} = -\not\!\epsilon$. From the normalization condition given in Eq. (\ref{Gno}), we obtain:
 \be
   \delta d_Q = \frac{e}{2 m_Q}.
 \en
Including the above results, we can write:
\be
   d&=& d_0 + \delta d_1^{(k)} + \delta d_1^{(g)} + \delta d_2,\non \\
   \tilde d_q &=& d + \frac{e_Q}{e_q} d_Q.
\en

\section{Numerical results and discussion}
For obtaining numerical results, we shall further assume the form of $\varphi(v\cdot p)$: (i) $\varphi(v\cdot p)$ is an analytic function apart from isolated singularities in the complex plane, and (ii) it vanishes as $|v\cdot p| \rightarrow \infty$. These two conditions
allow us to evaluate the $p^0-$ (or $p^--$) integrations in Eq. (\ref{Gno}) by Cauchy's Theorem. Thus, we take:
 \be
 \varphi_n(v\cdot p)=\frac{1}{(v\cdot p+\omega-i\varepsilon)^n} \qquad (n={\rm integer}),
 \en
which was used in a previous work \cite{CH14}.
There are some parameters ($m_s, m_Q, \omega, \alpha_s$) in this covariant model, and we follow the strategy described below to fix them. In a quark model, flavor $SU(3)$ symmetry is broken because the strange quark mass $(m_s)$ is quite different from the up or down quark mass $(m_{u,d})$. However, the size of the difference,
 \be
   \delta m_q = m_s - m_{u,d},
 \en
is not accurately known. For current quark masses, the value of $\delta m_q$ was quoted as $\delta m_q(\mu=1~\textrm{GeV})\simeq 190$ MeV \cite{1G} and $\delta m_q(\mu=2~\textrm{GeV})\simeq 90$ MeV \cite{PDG14} in the different renormalization scales. On the other hand, for constituent quarks in a relativistic quark model, one typically gets \cite{TT}:
 \be
   \delta m_q \simeq 140 \sim 200~\textrm{MeV}.\label{deltamq}
 \en
Because $\delta m_q$ is an important parameter in our calculations, using the variant values in the above range will leading to the quite different results. This will slash our predictive ability. Here we use the constraint that $\bar \Lambda_s$ is independent of the heavy quark mass to obtain the value of $\delta m_q$. In other words, $\delta m_q$ is no longer a free parameter. The processes are as follows: we first quote the value $m_{u,d}=0.245$ GeV from the previous work and try the initial value with $\delta m_q = 140$ MeV. Subsequently, we take the charm quark mass and the quark-gluon coupling to be the same as that for the non-strange charm meson
\cite{CH14}, and choose an $\omega$ to calculate
$\lambda_1^{(k)}$, $\lambda_1^{(g)}$, and $\lambda_2$ from Eqs. (\ref{lamonek}) $-$
(\ref{lamtwo}). Using Eq. (\ref{masslp}), the value of $\omega$ can be adjusted to fit the hyperfine
mass splitting \cite{PDG14}
\be
   \Delta M_{D^*_s D_s} &=& 143.8 \pm 0.4 ~\textrm{MeV}. \label{hyDs}
\en
After fixing $\omega$, we take the bottom quark mass and the quark-gluon coupling to be the same as that for the non-strange bottom meson
\cite{CH14} to estimate the other hyperfine
mass splitting, $\Delta M_{B^*_s B_s} $. In addition, using Eq. (\ref{MMM}), we can determine two values of $\bar \Lambda_s$ for both the charm and bottom sectors. Because $\bar \Lambda_s$ is independent of the heavy quark mass, the above processes
are repeated by fine-tuning the value of $\delta m_q$ until the two values of $\bar \Lambda_s$ are the same. Finally, the decay constant in the heavy quark limit, $\bar f_{M_s}$, can also be evaluated in terms of Eq. (\ref{fbm}). These results are listed in Tables \ref{tab:1} and \ref{tab:2}.
\begin{table}[htb]
\begin{center}
\begin{tabular}{|c|c|c|c|c|c|c|c|c|} \hline
$n$ & $\delta m_q$ (GeV) & $m_Q$ (GeV)& $\alpha_s$ & $\omega$ (GeV) & $\lambda_2$ (GeV$^2$)& $\lambda_1$ (GeV$^2$) & $\bar \Lambda_s$ (GeV) & $\bar f_{M_s}$ (GeV$^{3/2}$)
\\ \hline
$8$ & $0.225$ & $1.73$ & $0.400$ & $1.19$ & $0.124$ & $-0.210$  &$0.290$ & $0.507$ \\ 
$10$ & $0.219$ & $1.72$ & $0.392$ & $1.78$ & $0.124$ & $-0.232$  & $0.285$ & $0.508$\\ 
$12$ & $0.215$ & $1.72$ & $0.387$ & $2.38$ & $0.124$ & $-0.244$  & $0.281$ & $0.508$\\ \hline  
\end{tabular}
\end{center}
\caption{$D_s$-meson parameters for $\varphi_{n}$.}
\label{tab:1}
\end{table}

\begin{table}[htb]
\begin{center}
\begin{tabular}{|c|c|c|c|c|c|c|c|c|} \hline
$n$ & $\delta m_q$ (GeV) & $m_Q$ (GeV)& $\alpha_s$ & $\omega$ (GeV) & $\lambda_2$ (GeV$^2$)& $\lambda_1$ (GeV$^2$) & $\Delta M_{B^*_s B_s}$ (MeV)  & $\bar \Lambda_s$ (GeV) 
\\ \hline
$8$ & $0.225$ & $5.09$ & $0.381$ & $1.19$ & $0.118$ & $-0.230$ & $46.3$ & $0.290$ \\
$10$ & $0.219$ & $5.09$ & $0.373$ & $1.78$ & $0.118$ & $-0.251$ & $46.3$ & $0.285$ \\
$12$ & $0.215$ & $5.09$ & $0.368$ & $2.38$ & $0.118$ & $-0.264$ & $46.3$ & $0.281$ \\ \hline 
\end{tabular}
\end{center}
\caption{$B_s$-meson parameters for $\varphi_{n}$. 
}
\label{tab:2}
\end{table}

First of all, we see that the choice of $\varphi_n$ $(n=8,10,12)$ makes very little difference. The value of $\delta m_q=215\sim 225$ MeV is close to the typical light-quark mass, Eq. (\ref{deltamq}), used in a relativistic formalism \cite{TT}. The hyperfine mass splitting $\Delta M_{B^*_s B_s} $ is consistent with the average data: $\Delta M_{B^*_s B_s}^{\textrm{ave}}=46.1\pm 1.5$ MeV. However, the value of $\alpha_s$ in $B_s$ meson seems to be rather larger than the one which is determined 
by the perturbative evolution equation (at the one-loop level in the $\overline{\textrm{MS}}$ scheme):
 \be
 \alpha_s(m_b^{\textrm{pole}})=\frac{\alpha_s(M_Z)}{1+\alpha_s(M_Z)\beta_0 \textrm{ln}[(m_b^{\textrm{pole}}/M_Z)^2]/(4\pi)}\simeq 0.22,
 \en
where $m_b^{\textrm{pole}}=4.89$ GeV, $M_Z=91.19$ GeV, $\beta_0=11-\frac{2}{3} N_f=11-\frac{8}{3}$ for $N_f=4$, and $\alpha_s(M_Z)=0.119$ from experimental fits. The reason is that \cite{NN} if the gluons which coupling to the heavy quarks are hard (i.e., the virtual momenta is of order of the heavy quark mass), they can resolve the nonlocality  of the propagator of the small component fields $H_v\equiv i\not\!\!\!D_\perp h_v/(iv\cdot D+2 m_Q-i\epsilon)$. Their effects are not taken into account in the naive operator product expansion which was used in the derivation of the effective Lagrangian in Eq. (\ref{O12}). Thus, HQET provides an appropriate description only at scales $\mu \ll m_b$, and the relevant $\alpha_s$ in the $b$-quark mesons will larger than $\alpha_s(m_b^{\textrm{pole}})$. As to the reduced mass, we compare $\bar \Lambda_s=0.281\sim0.290$ GeV with that of the non-strange heavy meson, $\bar \Lambda=0.202\sim0.210$ GeV \cite{CH14}, and find that the residual mass difference is only about $80$ MeV, in contrast to $\delta m_q = 215 \sim 225$ MeV. This can be understood as follows. Due to its heavier mass, the strange quark is more tightly bound than an up or down quark; thus, part of the mass difference $\delta m_q$ is compensated for by a larger binding energy of the $(Q\bar s)$-system. We can then obtain the predicted meson decay constant $f_{M_s}$ by using $f_{M_s}=\bar f_{M_s}/\sqrt{M_{M_s}}$
and the ratio $f_{M_s}/f_M$:
 \be
   f_{M_s} \simeq 219~\textrm{MeV},~~~\frac{f_{M_s}}{f_M} = 1.13\pm 0.05,
 \en
where the value $f_M\simeq f_B=194 \pm 9$ MeV (an average of the results \cite{lattice1,lattice2} in lattice QCD) is chosen. For comparison, the QCD sum rules results of \cite{QSR1}
 \be
   f_{B_s} = 242^{+17}_{-12}~\textrm{MeV},~~~\frac{f_{B_s}}{f_B} = 1.17^{+0.03}_{-0.04},
 \en
and \cite{QSR12,QSR2}
 \be
   f_{B_s} = 225.6 \pm 18.3 \pm 3~\textrm{MeV},~~~\frac{f_{B_s}}{f_B} = 1.184 \pm 0.023 \pm 0.007,
 \en
and the lattice QCD calculation results \cite{lQCD3} of
\be
   f_{B_s} = 224(5)~\textrm{MeV},~~~\frac{f_{B_s}}{f_B} = 1.205(7),
\en
are shown here. In Tables I and II, the kinetic and chromomagnetic expectation values, $\lambda_1$ and $\lambda_2$, are the heavy-strange meson parameters which were defined in some papers as $\mu_\pi^2$ and $\mu_G^2$, respectively. The relations between them are \cite{mulambda}:
 \be
 \lambda_1=-\mu_\pi^2,~~~~~\lambda_2=\mu_G^2/3. \label{mulambda12}
 \en
Although these heavy-strange meson parameters are not found in the other theoretical calculations, here we show the relevant parameters of $B$ meson ($n=10$) which obtained from the previous work \cite{CH14}: $\lambda_1=-0.162~\textrm{GeV}^2, \lambda_2=0.117~\textrm{GeV}^2$. Compared with the recent result \cite{AGH} which comes from the inclusive decays with $m_b^{\textrm{kin}}=4.553~\textrm{GeV}$ (Eq. (\ref{mulambda12}) is used): $\lambda_1=-0.465~\textrm{GeV}^2, \lambda_2=0.111~ \textrm{GeV}^2$, and we find that $\lambda_2$ is consistent with ours because of a bound from the $B$ hyperfine splitting, but otherwise $\lambda_1$ is rather different from ours.

Next, the $1/m_Q$ corrections to $f_s$ and $ d_s$ are listed in Table \ref{tab:3}. 
\begin{table}[htb]
\begin{center}
\begin{tabular}{|c|cccc|cccc|cccc|} \hline
& \multicolumn{8}{c} {$D^*_sD_s$} \vline & \multicolumn{4}{c} {$B^*_sB_s$} \vline \\ \hline
$n$ & $f_{s0}$ & $\delta f_{s1}$ & $\delta f_{s2}$ & $f_s$ & $d_{s0}$ & $\delta d_{s1}$ & $\delta d_{s2}$ & $d_s$ & $ d_{s0}$ & $\delta d_{s1}$ & $\delta d_{s2}$ & $d_s$   \\ \hline
$8$ & $-3.29$ & $0.292$ & $ -0.190$& $-3.18$ & $0.529$ &$-0.0382$& $0.0307$& $0.522$ &
$0.529$  &$-0.0146$ & $0.00988$ & $0.525$ \\ \hline
$10$ & $-3.10$ & $0.303$& $-0.184$ & $-2.98$ & $0.505$ & $-0.0399$& $0.0299$ & $0.495$ &
$0.505$ & $-0.0150$& $0.00962$ & $0.500$ \\  \hline
$12$ & $-2.99$ & $0.308$ & $-0.181$ & $-2.86$ & $0.491$ & $-0.0408$& $0.0294$ & $0.479$ &
$0.491$ &$-0.0153$ & $0.00948$ & $0.485$ \\  \hline
\end{tabular}
\end{center}
\caption{$1/m_Q$ corrections to $f$ and $d$. }
\label{tab:3}
\end{table}
In order to do a comparison for the $1/m_Q$ effects, we introduce an effective gluon mass of $m_g \simeq \Lambda_{\textrm{QCD}} \simeq 300$ MeV. For $D^*_s D_s$ mesons,
\be
   \left(\frac{\delta f_{s2}}{f_s}\right) = 5.97~\%,~~~~~
   \left(\frac{\delta d_{s2}}{d_s}\right) = 5.88~\%, \non
\en
are consistent with the rough estimate of
\be
   \alpha_s \frac{\Lambda_{\textrm{QCD}}}{m_c} \sim (6\sim 7)~\%,
\en
for $\alpha_s = 0.4$. For $B^*_s B_s$ mesons, since $m_b/m_c \simeq 3$, consequently, the HQS violating effect and the rough estimation are both smaller than those for the $D^*_sD_s$ system by approximately a factor of $3$. Additionally, as compared with that of the non-strange charm mesons \cite{CH14}, we see that the $SU(3)$ breaking is severe for $f_s$, but less so for $d_s$ (see Table \ref{tab:4}).
\begin{table}[htb]
\begin{center}
\begin{tabular}{|c|c|c|c|c|c|} \hline
$m_{u,d}$(GeV) & $f$ & $d$(GeV)$^{-1}$ & $\delta m_q$(GeV) & $f_s$ & $d_s$(GeV)$^{-1}$  \\ \hline \hline
$0.245$ & $-1.13$ & $0.361$  & $0.225$ & $-3.18$ & $0.522$  \\ \hline
\end{tabular}
\end{center}
\caption{$SU(3)$ symmetry breakings to $f$ and $d$ for $n=8$.}
\label{tab:4}
\end{table}
The reason for this can be traced back to the fact that the strong coupling constant is sensitively dependent on $\bar \Lambda_s$, but the magnetic coupling constant is sensitively dependent on both $\bar \Lambda_s$ and $m_s$ in this model. The details are as follows. From Eqs. (\ref{gg}) and (\ref{D11}), both the strong and magnetic coupling constants are dependent on $\bar \Lambda_s$. As the power of $|\vec{p}|$ in the wave function $\varphi$ must be smaller than $-\frac{2}{3}$ (see Eq. (\ref{Gno})), the strong and magnetic coupling constants satisfy a simple relation, $d_{s0}=\frac{-e}{2 m_s} \frac{f_{s0}}{2}$, which is similar to the result in Appendix A of Ref. \cite{CH14}. Combined with the estimation that the total $1/m_Q$ correction is about $(3\sim4) \%$ for the $D^*_s D_s$ mesons, we obtain an approximate equation, $d_{s}\simeq\frac{-e}{2 m_s} \frac{f_{s}}{2}$. Therefore, the $SU(3)$ breaking of $d_s$ has been reduced by the factor $m_s$ in the denominator.

The study of $SU(3)$ breaking in chiral perturbation theory follows a different route, in which $SU(3)$ symmetry is assumed at the tree level and symmetry breaking effects are induced via meson loops (see \cite{Randall} for details). Thus, from Figure \ref{su3} (a) we have:
\begin{figure}
 \includegraphics*[width=4.5in]{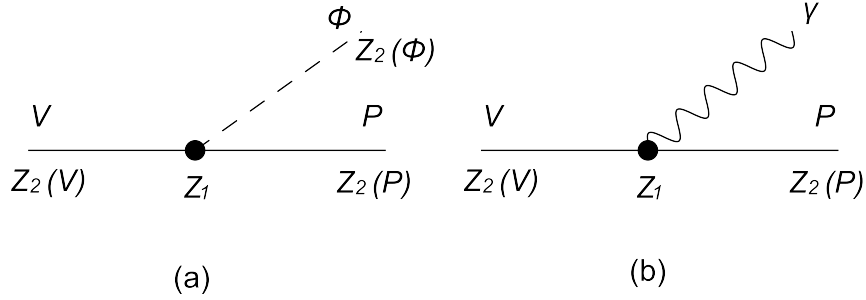}
 \caption{Renormalizations of (a) the strong coupling constant $f$ and (b) the $VP\gamma$ coupling constant $d$ in chiral perturbation theory.}
  \label{su3}
\end{figure}
\be
   f = \frac{\sqrt{Z_2(V) Z_2(P) Z_2(\phi)}}{Z_1(VP\phi)} f^0 \label{2112}
\en
where $Z_1$ and $Z_2$ are, respectively, the wave function and vertex renormalization constants, $\phi$ denotes a Goldstone boson, and $f^0$ is the unrenormalized coupling constant. The $Z$s have all been evaluated in \cite{HYC4}. Putting in the numbers in (\ref{2112}), we obtain:
 \be
   f = 1.33,~~~f_s = 1.47, \non
 \en
for $f^0=0.52$, which fits to the experimental data for non-strange mesons \cite{HYC4}. Thus, we see that in chiral perturbation theory, $SU(3)$ breaking in the strong coupling constant is not large, with:
 \be
   \frac{f_s- f}{f}\Bigg |_{\textrm{chiral}} \sim 0.10 \label{2113}
 \en
This is very different from what we found in the covariant model. As for the radiative decay constants in chiral perturbation theory, we have (see Figure \ref{su3} (b)):
 \be
   d = \frac{\sqrt{Z_2(V) Z_2(P)}}{Z_1(VP\gamma)} d^0 \label{AB}
 \en
where $d_0$ is the unrenormalized transition magnetic moment, and $d^0 = 0.394~\textrm{GeV}^{-1}$ is obtained from fitting to the branching ratios of $D^* \to D\gamma$ \cite{HYC4}. Putting the numbers in (\ref{AB}), we obtain $d = 0.436~\textrm{GeV}^{-1}, d_s = 0.575~\textrm{GeV}^{-1}$ and
 \be
    \frac{d_s - d}{ d}\Bigg |_{\textrm{chiral}} = 0.319.\non
 \en
The latter one is close to that of our model:
 \be
   \frac{d_s - d}{d} = 0.446.\non
 \en

\begin{figure}
 \includegraphics*[width=4.5in]{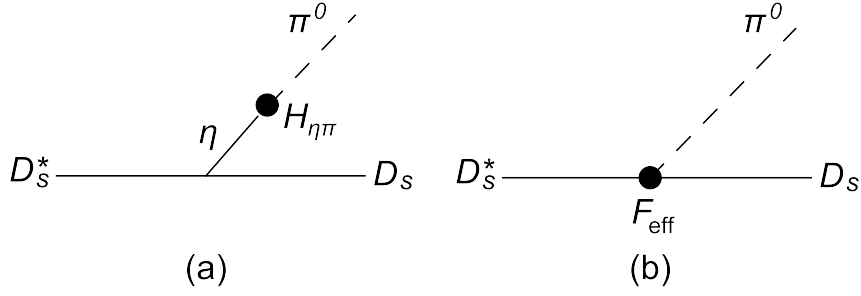}
 \caption{$D^*_s \to D_s + \pi^0$ via $\eta \pi$-mixing mechanism.}
  \label{su2}
\end{figure}

Finally, we consider the decay widths $\Gamma(D^*_s \to D_s \pi^0)$, $\Gamma(D^*_s \to D_s \gamma)$, $\Gamma(B^*_s \to B_s \gamma)$ and the ratio:
 \be
   r_s = \frac{\Gamma(D^*_s \to D_s \pi^0)}{\Gamma(D^*_s \to D_s \gamma)}, \label{2121}
 \en
which is known to be $r_s = 0.062 \pm 0.008$ experimentally \cite{PDG14}. Note that the decay mode
 \be
   D^*_s \to D_s + \pi^0
 \en
violates isospin or $SU(2)$ symmetry, and it must proceed via $\eta-\pi$ mixing in the leading order \cite{ChoWise,Ste}, as depicted in Figure \ref{su2} (a),
where $H_{\eta\pi}=\langle\pi^0 |H_{\textrm{em}}|\eta\rangle$ is the $\eta$-$\pi$ mixing vertex. Figure \ref{su2} (a) can be replaced by Figure \ref{su2} (b) with an effective $D^*_s D_s \pi$ coupling constant:
 \be
   F_{\textrm{eff}} = \frac{-2}{\sqrt{3}} f_s \frac{H_{\eta\pi}}{m^2_\pi - m^2_\eta}.
 \en
The strength of the $\eta$-$\pi$ mixing interaction, $H_{\eta\pi}$, can be calculated in various models \cite{GL}. Here we chose to utilize the experimental rates of $\eta \to 3\pi^0$. From the data of Particle Data Group \cite{PDG14}, we can use $\Gamma (\eta\to \textrm{all})=1.31\pm0.05$ keV and $\textrm{Br}(\eta \to 3\pi^0)=(32.68\pm 0.23)\%$ to obtain:
\be
\Gamma(\eta \to 3\pi^0)=0.428\pm 0.019~~\textrm{keV}.\label{width}
\en
As to the amplitude of $\eta \to 3\pi^0$, a fit of the data in Ref. \cite{PDG14} shows $M_{\eta\to 3\pi^0}$ to be essentially constant over phase space: $|M_{\eta\to 3\pi^0}|^2=M^2_0(1+2\alpha z)$, where $z$ is the square of the relative distance to the center of the Dalitz plot and $\alpha=-0.0315\pm 0.0015$. Then the three-body phase space integral for constant amplitude was estimated in Ref. \cite{ref12}:
\be
\Gamma(\eta\to 3\pi^0)\simeq 0.827~|M_{\eta\to 3\pi^0}|^2~~\textrm{keV}. \label{constantM}
\en
Combining Eqs. (\ref{width}) and (\ref{constantM}), we obtain the constant amplitude:
\be
|M_{\eta\to 3\pi^0}|=0.719\pm0.033.\label{Mexp}
\en
On the other hand, from the current-algebra PCAC \cite{PCAC}, the total amplitude of $\eta \to 3\pi^0$ is summing the three cyclic permutations of Figure \ref{cyclic} \cite{ref13,CMS}:
\begin{figure}
 \includegraphics*[width=5in]{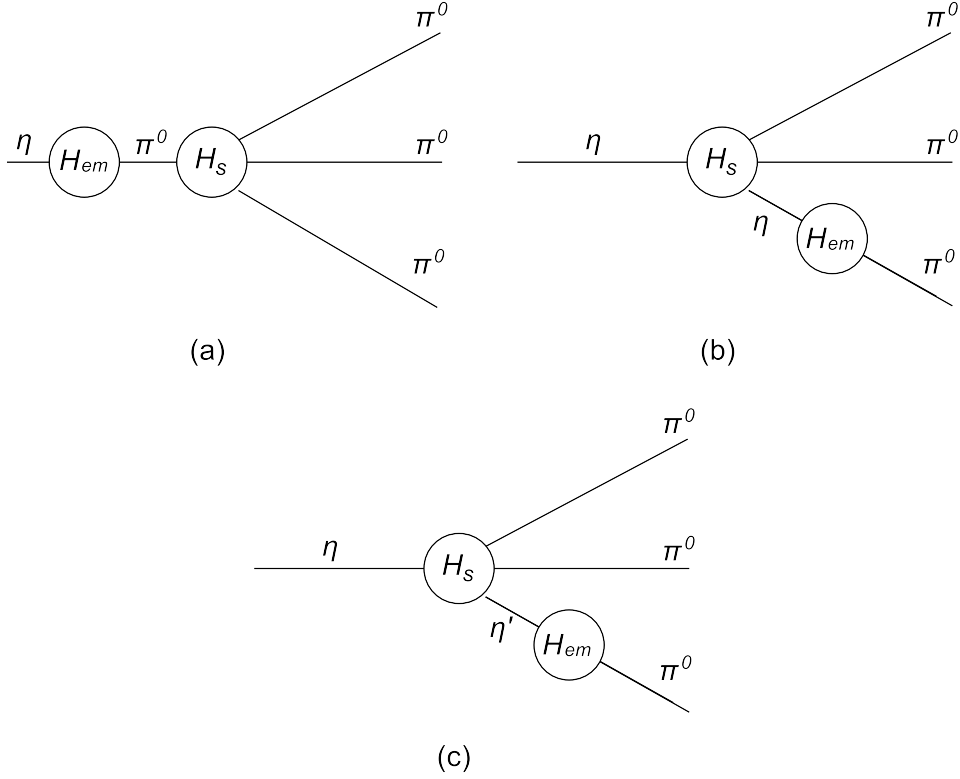}
 \caption{One of three cyclic amplitudes in $\eta \to 3\pi^0$.}
  \label{cyclic}
\end{figure}
\be
M_{\eta\to3\pi^0} = \frac{\langle\pi^0|H_{\textrm{em}}|\eta\rangle}{m^2_\eta-m^2_\pi}\bigg[2\frac{m^2_\eta}{f^2_\pi}
-M^{\textrm{st}}(\eta\to\eta_\pi2\pi^0)-M^{\textrm{st}}(\eta\to\eta'_\pi 2\pi^0)\frac{\langle\pi^0|H_{\textrm{em}}|\eta'\rangle}{\langle\pi^0|H_{\textrm{em}}|\eta\rangle}
\frac{m^2_\eta-m^2_\pi }{m^2_{\eta'}-m^2_\pi}\bigg]\non \\\label{etato3pi},
\en
where the first, second, and third terms of Eq. (\ref{etato3pi}) correspond to $H_s$ in Figs. \ref{cyclic}(a), (b), and (c), respectively. The former is the Weinberg $\pi\pi \to \pi\pi$ \cite{Weinberg} strong amplitude extrapolated to the $\eta$ mass shell consistent with four-momentum conservation, and the latter is the strong amplitude $M^{\textrm{st}}(\eta\to\eta^{(_{'})}_\pi2\pi^0)$ of the $\eta^{(_{'})}$ pole which extrapolated from the $\eta^{(_{'})}$ mass to the $\pi$ mass. Summing the cyclic permutations and considering the $\eta\eta'$ mixing, the authors of Ref. \cite{CMS} obtained: $M^{\textrm{st}}(\eta\to\eta_\pi2\pi^0)=6 \cos^2\phi~m^2_\pi/f^2_\pi$, where $\phi$ is the mixing angle:
\be
|\eta\rangle &=& \cos\phi|{\bar n}n\rangle-\sin\phi|{\bar s} s\rangle,\non \\
|\eta'\rangle &=& \sin\phi|{\bar n}n\rangle+\cos\phi|{\bar s} s\rangle,
\en
with $|{\bar n}n\rangle=(|{\bar u}u\rangle+|{\bar d}d\rangle)/\sqrt{2}$, and the third term of Eq. (\ref{etato3pi}) is negligible. The mixing angle can be determined by a theory of particle mixing \cite{ref8}:
\be
\tan^2\phi =\frac{(m_{\eta'}^2-2 m^2_K+m^2_\pi)(m^2_\eta-m^2_\pi)}{(2m_K^2-2 m^2_\pi-m^2_\eta)(m^2_{\eta'}-m^2_\pi)},
\en
from which one obtains: $\phi=41.5^\circ$.
Combining Eqs. (\ref{Mexp}) and (\ref{etato3pi}) with the pion decay constant $f_\pi=130.41\pm 0.20$ MeV and the relevant meson masses \cite{PDG14}, we obtain:
\be
   H_{\eta\pi} = -6400 \pm 310~\textrm{MeV}^2.
\en
This value is different from that of Ref. \cite{CMS}: $H_{\eta\pi} = -5900 \pm 600~\textrm{MeV}^2$, the chief reason being that the new experimental data in Eq. (\ref{width}) is used. Then, using the $f_s$ and $d_s$ obtained in Table \ref{tab:3}, we can predict the ratio $r_s$ and the relevant decay widths in Table \ref{tab:5}.
\begin{table}
\begin{center}
\begin{tabular}{|c|c|c|c|c|} \hline
$n$ & $D^*_s \to D_s \pi^0$ & $D^*_s \to D_s \gamma$ & $r_s$ & $B^*_s \to B_s \gamma$ \\ \hline  \hline
$8$ & $0.277^{+0.028}_{-0.026}(7.29^{+0.67}_{-0.65}\%)$ & $3.53(92.7\mp0.7 \%)$ & $0.0786^{+0.0079}_{-0.0075}$ & $0.407$ \\ \hline
$10$ & $0.243^{+0.024}_{-0.023}(7.48^{+0.69}_{-0.67}\%)$ & $3.00(92.5\mp0.7 \%)$ &$0.0809^{+0.0081}_{-0.0077}$ & $0.371$ \\  \hline
$12$ & $0.224^{+0.022}_{-0.022}(7.60^{+0.70}_{-0.67}\%)$ & $2.72(92.4\mp0.7 \%)$ &$0.0823^{+0.0083}_{-0.0079}$ & $0.351$\\ \hline
\end{tabular}
\end{center}
\caption{Predicted decay rates (in unit of keV), the branching ratios (in parentheses) and $r_s$ for $\varphi_n$.}
\label{tab:5}
\end{table}
For the different $\varphi_n$, the deviations of the decay rate are about $5\sim6$ times larger than those of $r_s$. The main reason is that the decay rate $\Gamma(D^*_s\to D_s\gamma)$, for example, can be simplified as:
\be
\Gamma(D^*_s\to D_s\gamma)\propto \bigg|-d_s+\frac{e}{m_c}\bigg|^2,\label{gammapro}
\en
where the minus sign comes from the charge of the $s$ quark. On the other hand, the ratio $r_s$ can be simplified as:
\be
r_s\propto \bigg|\frac{d_s m_s}{-d_s+\frac{e}{m_c}}\bigg|^2,\label{rspro}
\en
where $d_s\simeq \frac{-e}{2 m_s}\frac{f_s}{2}$ is again applied. In other words, the deviations of $r_s$ for the different $\varphi_n$ are suppressed strongly because $d_s$ is approximately proportional to $f_s$. Finally, we list the predicted decay rates and branch ratios within this work ($n=8$) and some theoretical models in Table \ref{tab:6}. For comparison, the experimental data are also included.
{\small
\begin{table}[htb]
\begin{center}
\begin{tabular}{|c|c|c|c|c|c|} \hline
 Reaction & $D^*_s \to D_s\pi^0$ & $D^*_s \to D_s\gamma$ & $D^*_s \to$ total & $r_s(\times 10^{-2})$ & $B^*_s \to B_s\gamma$  \\ \hline
Exp.\cite{PDG14} & $(5.9 \pm 0.7 ~\%)$ & $(94.2\pm 0.7~\%)$ & $<1900$ & $6.2\pm 0.8$ &  \\
This work & $0.277^{+0.028}_{-0.026}(7.3\pm 0.7\%)$ & $3.53(92.7 \mp 0.7 \%)$ & $3.56\pm0.03$ & $7.86^{+0.79}_{-0.75}$ & $0.407$ \\ 
$\chi$PT \cite{HYC4}$^\dagger$ &  & $4.5$ & & &  \\
LFQM \cite{Choi}$^\ddag$ & & $0.18\pm 0.01$& & & $0.068\pm 0.017$ \\
RQM \cite{Goity}$^\sharp$ & $0.0197\pm0.0070$~(input)  & $0.321^{+0.009}_{-0.008}$ & $0.341$ & & $0.136\pm 0.012$ \\
QCDSR \cite{QCDSR} &  & $0.59\pm 0.15$ & & & \\
NJLM \cite{NJLM} &  & $0.09$ &  & & $0.10$\\
LQCD \cite{latticeQCD} & $0.0040$~(input)& $0.066\pm 0.026$ & $0.070\pm 0.028$ & &  \\
MIT \cite{MIT} & & & & & $0.0510$ \\
NRQM \cite{Kamal} & & $0.21$ & & &  \\
NRQM \cite{FM}$^\natural$ & & $0.40$ & & & $0.18$ \\ \hline
\end{tabular}
\end{center}
\caption{Predicted decay rates (in units of keV) and branch ratios (in parentheses) of some models. For comparison, the experimental branching ratios are given in the first row. ($\chi$PT: chiral perturbation, LFQM: light-front quark model, RQM: relativistic quark model, QCDSR: QCD sum rules, NJLM: Nambu-Jona-Lasinio model, LQCD: lattice QCD, MIT: MIT bag model, NRQM: non-relativistic quark model.) $\dagger$: The value for $g=0.52$, $\beta=2.6$ GeV$^{-1}$, and $m_c=1.6$ GeV. $\ddag$: The values correspond to a linear model. $\sharp$: The value for $\kappa^q=0.55$. $\natural$: The values correspond to $(a)$ model.}
\label{tab:6}
\end{table}}
We can find that, first, our branching ratios and $r_s$ are close to the experimental data. Second, there are many theoretical calculations of $\Gamma(D^*_s \to D_s\gamma)$ in the literature. However, except for the result of $\chi$PT, our result is much larger than that of the other theoretical methods. In fact, the estimations of $\Gamma(D^*_s \to D_s\gamma)$ are also quite different among the other theoretical groups. A similar situation exists for $\Gamma(B^*_s \to B_s\gamma)$. Third, in contrast to $\Gamma(D^*_s \to D_s\gamma)$, there are few computations for $\Gamma(D^*_s \to D_s\pi)$ violating the isospin symmetry. The relevant results of Refs. \cite{Goity,latticeQCD} come from taking the experimental branching ratio as the input. Then, for $\Gamma(D^*_s \to D_s\pi)$ as well as $r_s$, we need to make further comparisons by means of more experiments and theoretical calculations. It is worth mentioning that, in our model, although the deviations of $r_s$ are smaller than those of the decay rate for the different $\varphi_n$ (because of $d_s \propto f_s$; see above), this does not mean that we can obtain almost the same $r_s$, no matter what, for example, the value of $\Gamma(D^*_s \to D_s\gamma)$ is. The average value of $\Gamma(D^*_s \to D_s\gamma)$ for the other theoretical calculations (except  $\chi$PT) is about one-tenth of ours. From Eqs. (\ref{gammapro}) and (\ref{rspro}), we find that if one adjusts the $d_s$ to reduce the value of $\Gamma(D^*_s \to D_s\gamma)$ to one-tenth, the value of $r_s$ will be enhanced to about $3\sim4$ times that of the experimental data. In other words, the fact that our $r_s$ is close to the experimental data gives us confidence in our results and the validity of our covariant framework.
\section{Conclusions}
Based on HQET, we have discussed the strong and radiative coupling constants of strange heavy mesons in $1/m_Q$ corrections and $SU(3)$ symmetry breakings. These effects were studied using a fully covariant model. The covariant model starts from HQET in the heavy quark limit and describes a heavy meson as a composite particle, consisting of a reduced heavy quark coupled with a brown muck of light degrees of freedom. It is formulated in an effective Lagrangian approach, so that it is fully covariant, and we used Feynman diagrammatic techniques to evaluate the various processes. 

The parameters of this model, $m_s$ and $\omega$, were chosen to fit the data of the hyperfine mass splitting, $\Delta M_{D^*_s D_s}$, and because the residual mass $\bar \Lambda_s$ is independent of the heavy quark mass. Then the other hyperfine mass splitting $\Delta M_{B^*_s B_s}$, $\bar \Lambda_s$ and the decay constant in the HQ limit can be calculated. Our $\Delta M_{B^*_s B_s}$ was consistent with the data. The residual mass difference between $\bar \Lambda_s$ and $\bar \Lambda$ was only about $80$ MeV, and obviously smaller than that between the $s$ and $u,d$ quarks. This is understood as follows. Due to its heavier mass, the strange quark is more tightly bound than an up or down quark; thus, part of the mass difference between the $s$ and $u,d$ quarks is compensated for by a larger binding energy of the $(Q\bar s)$-system. The $SU(3)$ symmetry breaking effect $f_{M_s}/f_{M}$ is close to the $f_{B_s}/f_B$ of the QCD sum rules \cite{QSR1,QSR12,QSR2} and the lattice QCD calculation \cite{lQCD3} results.

The $1/m_Q$ corrections of $f_s$ and $d_s$ are consistent with $\alpha_s \Lambda_{\textrm{QCD}}/m_Q$ for both the $D^*_s D_s$ and the $B^*_s B_s$ systems. In the charmed meson sector, the HQS violating effects are larger by approximately a factor of $3$ because of $m_b/m_c \simeq 3$. The $SU(3)$ symmetry violating percentage of $f_s$, at about $180\%$, is obviously larger than that of the $d_s$ at $45\%$. The reason for this was that, due to an approximate equation, $d_s\simeq \frac{e}{2m_s}\frac{f_s}{2}$, the $SU(3)$ breaking of the $d_s$ was reduced by the factor $m_s$ in the denominator. For comparison, we estimated the $SU(3)$ symmetry violating percentages in chiral perturbation theory \cite{HYC4} and obtained about $10\%$ and $32\%$ for $f_s$ and $d_s$, respectively.

In order to calculate the decay rate of $D^*_s\to D_s \pi$, which violates isospin or $SU(2)$ symmetry, we used the new data of $\Gamma(\eta \to 3\pi^0)$ to estimate the $\eta-\pi$ mixing vertex: $H_{\eta\pi}=-6400\pm 310$ MeV$^2$. Combining the coupling constants $f_s$, $d_s$, and $H_{\eta\pi}$, we studied $\Gamma(D^*_s \to D_s \pi^0)$, $\Gamma(D^*_s \to D_s \gamma)$, $\Gamma(B^*_s \to B_s \gamma)$, and the ratio $r_s$. For the different $\varphi_n$, the deviations of $\Gamma(D^*_s \to D_s \pi^0)$ and $\Gamma(D^*_s \to D_s \gamma)$ were about $5\sim6$ times larger than those of $r_s$. The main reason was $d_s\simeq \frac{e}{2m_s}\frac{f_s}{2}$, again. In other words, the deviations of $r_s$ are suppressed strongly because $d_s$ is approximately proportional to $f_s$.

Finally, we compared our results with the experimental data and the other theoretical calculations in Table VI. Our branching ratios of $\Gamma(D^*_s \to D_s \pi^0)$, $\Gamma(D^*_s \to D_s \gamma)$, and $r_s$ were close to the experimental data. However, our predictive decay widths of $\Gamma(D^*_s \to D_s \pi^0)$, $D^*_s \to D_s \gamma$, and $B^*_s \to B_s \gamma$ were much larger than those of the other theoretical groups except for $\chi$PT. Because another computational $r_s$ is not found in the literature, we tried to reduce our $\Gamma(D^*_s \to D_s \gamma)$ to the average value of the other theoretical calculations by directly adjusting $d_s$, and find that our $r_s$ is enhanced to about $3 \sim 4$ times of the experimental data. In other words, the fact that our $r_s$ is close to the experimental data gives us confidence not only in the validity of our covariant framework, but also in our predictions about the decay widths of $D^*_s \to D_s \pi$, $D^*_s \to D_s \gamma$, and $B^*_s \to B_s \gamma$. Then more experiments about the above decay widths are needed.


{\bf Acknowledgements}\\
This work was supported in part by the National Science
Council of the Republic of China under Grant No.
NSC 102-2112-M-017-001-MY3.



\end{document}